\UseRawInputEncoding
\documentclass[twocolumn,secnumarabic,amssymb,nobibnotes,aps,prl]{revtex4-1}

\usepackage{graphicx}% Include figure files
\usepackage{dcolumn}% Align table columns on decimal point
\usepackage{bm}% bold math
\usepackage{multirow}
\usepackage{mathptmx}%Ó¢ÎÄ×Ö£¨°üÀ¨Êýѧ¹«Ê½ÖеÄÓ¢ÎÄ£©×ÖÌåΪNew Times Roman
\usepackage{amsmath}%¸ß¼¶Êýѧģʽºê°ü
\usepackage{amssymb}%¸ß¼¶Êýѧ·ûºÅºê°ü
\usepackage{bibentry,natbib}
\usepackage{setspace}
\begin{document}
\title{Controlling core-hole lifetime through an x-ray planar cavity}%

\author{Xin-Chao Huang,$^{1}$ Xiang-Jin Kong,$^{2}$ Tian-Jun Li,$^{1}$ Zi-Ru Ma,$^{1}$ Hong-Chang Wang,$^{3}$ \\Gen-Chang Liu,$^{4}$ Zhan-Shan Wang,$^{4}$ Wen-Bin Li,$^{4}$}%

\email{wbli@tongji.edu.cn}
\author{Lin-Fan Zhu,$^{1}$}%
\email{lfzhu@ustc.edu.cn}

\affiliation{$^1$Hefei National Laboratory for Physical Sciences at Microscale and Department of Modern Physics, University of Science and Technology of China, Hefei, Anhui 230026, People's Republic of China\\
$^2$Department of Physics, National University of Defense Technology, Changsha, Hunan 410073, People's Republic of China\\
$^3$Diamond Light Source, Harwell Science and Innovation Campus, Didcot, Oxfordshire, OX11 0DE, UK\\
$^4$MOE Key Laboratory of Advanced Micro-Structured Materials, Institute of Precision Optical Engineering (IPOE), School of Physics science and Engineering, Tongji University, Shanghai 200092, People's Republic of China}

\begin{abstract}
It has long been believed that core-hole lifetime (CHL) of an atom is an intrinsic physical property, and controlling it is significant yet is very hard. Here, CHL of the 2$p$ state of W atom is manipulated experimentally through adjusting the emission rate of a resonant fluorescence channel with the assistance of an x-ray thin-film planar cavity. The emission rate is accelerated by a factor linearly proportional to the cavity field amplitude, that can be directly controlled by choosing different cavity modes or changing the angle offset in experiment. This experimental observation is in good agreement with theoretical predictions. It is found that the manipulated resonant fluorescence channel even can dominate the CHL. The controllable CHL realized here will facilitate the nonlinear investigations and modern x-ray scattering techniques in hard x-ray region.

\textrm{PACS: 32.80.-t, 32.80.Qk, 42.50.Ct, 32.30.Rj, 78.70.Ck.}
\end{abstract}

\date{\today}%

\maketitle

%\tableofcontents

%\section{Introduction}
The particularity of an inner-shell excitation or ionization is to produce a core vacancy, which has a finite lifetime, i.e., the so-called core-hole lifetime (CHL), and then it decays into lower-lying states. There are two main relaxation pathways, i.e., radiative (fluorescence) and non-radiative (Auger decay or autoionization) channels, and the CHL is determined by the total decay rate of all relaxation channels. Normally, Auger effect dominates the decay routes of K shell for low-Z atoms~\cite{Auger1925} and L and M shells for higher-Z atoms~\cite{Krause1979}, so the CHL is sometimes called Auger lifetime. The CHL has long been considered as an intrinsic factor and controlling it is very difficult because the relaxation channels are hard to be manipulated with common methods.

Nevertheless, an adjustable CHL is strongly desired, since CHL changes are useful to detect ultrafast dynamics. An adjustable CHL is needed to give a deep insight to nonlinear light-matter interaction with the advent of x-ray free electron laser (XFEL), since the ratio of CHL to XFEL pulse width does matter for multiphoton ionization~\cite{Fukuzawa2013}, two-photon absorption~\cite{Tamasaku2018}, population inversion~\cite{yoneda2015atomic} and stimulated emission~\cite{PhysRevLett.117.027401,PhysRevLett.121.137403}. CHL is also a key factor in resonant x-ray scattering (RXS) process~\cite{gel1999resonant}, where the dynamics of the core-excited state is controlled by the duration time which is determined by both energy detuning and CHL~\cite{PhysRevA.59.380}. Because of a lack of an efficient method to manipulate CHL experimentally, the controlling schemes for duration time were based on the energy detuning up to now~\cite{PhysRevLett.77.5035,PhysRevA.69.022707,PhysRevX.3.011017,feifel2011core,morin2012ultrafast,miron2011handbook}.
The dynamics of the core-excited state determines the application range for RXS techniques, e.g., resonant inelastic x-ray scattering (RIXS)~\cite{RevModPhys.83.705}. Since Coulomb interaction between core-hole and valence electrons only exists during the existence of core-excited state, the relative timescale between CHL and elementary excitations governs the effectiveness of indirect RXIS~\cite{PhysRevB.75.115118,PhysRevLett.103.117003,dean2012spin}, especially for charge and magnon excitations~\cite{van2007theory,PhysRevLett.103.117003,PhysRevLett.105.167404,PhysRevX.6.021020,tohyama2018spectral}. In time-resolved RIXS (tr-RIXS), CHL also needs to be flexibly adjusted for pursuing higher time resolution~\cite{dean2016ultrafast,wang2018theoretical,PhysRevB.99.104306,buzzi2018probing}. Therefore, a controllable CHL will be very useful thus is strongly wished for, from both fundamental and application perspectives.

Because CHL is determined by the total decay rate of all relaxation channels, controlling CHL means manipulatable decay channels, at least one of them, which is a challenging task. Stimulated emission channel could be opened by intense and short x-ray pulses to accelerate CHL~\cite{PhysRevLett.117.027401,PhysRevLett.121.137403}, while such scheme can only be implemented in XFEL. The present work proposes another scheme that controls the spontaneous emission channel. R. Feynman once said, the theory behind chemistry is quantum electrodynamics (QED)~\cite{ralf2004}, indicating that the spontaneous emission rate of atom depends on the environment (photonic density of states). A cavity is such an outstanding system to robustly structure environment and modify the spontaneous emission rate in visible wavelength regime~\cite{PhysRevA.51.2545,RevModPhys.73.565}, as known cavity-QED. With the dramatic progress of new generation x-ray source and thin-film technology, cavity-QED effect in hard x-ray range was demonstrated by the laboratory of thin film planar cavity with nuclear ensembles~\cite{Rohlsberger1248,rohlsberger2012electromagnetically,PhysRevLett.111.073601,PhysRevLett.114.207401,PhysRevLett.114.203601,haber2017rabi} or electronic resonance~\cite{Haber2019}, which breeds the new field of x-ray quantum optics~\cite{adams2013x}.

In this work, a controllable CHL for 2$p$ state of W atom is realized through adjusting the emission rate of a resonant fluorescence channel with the assistance of an x-ray thin-film planar cavity. WSi$_2$ has a remarkable white line around the L$_{\textrm{III}}$ edge of W, which is a resonant channel and generally known to be associated with an atomic-like electric dipole allowed transition, from an inner shell 2$p$ to an unoccupied level 5$d$~\cite{Haber2019,PhysRevB.15.738,PhysRevB.19.679}. Inside the cavity, the emission rate of the resonant channel depends on the photonic density of states where the atom locates, which can be modified by the cavity field amplitude in experiment. Because the thin-film planar cavity can only enhance the photonic density of states, but not suppress it, only CHL shortening is realized in the present experiment. As long as the cavity effect is strong enough, the total decay rate will have measurable changes and lead to an controllable CHL.

%\section{Model}
\begin{figure}[htbp]
\centering
\includegraphics[width=0.45\textwidth]{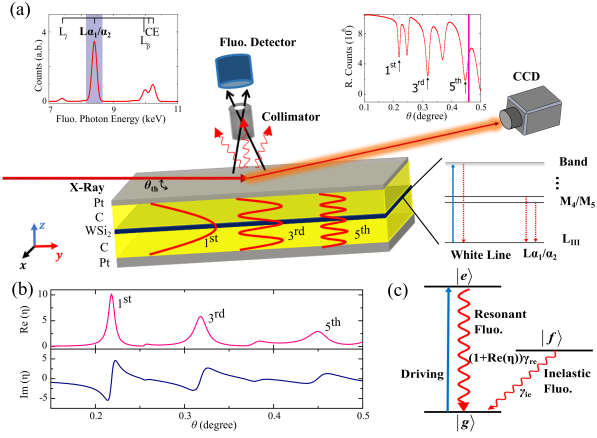}
\renewcommand{\figurename}{Fig.}
\caption{\label{Fig.1}The schematic for controlling core-hole lifetime. (a) Cavity sample and measurement setup. The cavity has a structure of Pt (2.1 nm)/C (18.4 nm)/WSi$_2$ (2.8 nm)/C (18.0 nm)/Pt (16.0 nm)/Si$_{100}$, and the middle-right inset shows the energy-level of L$_{\textrm{III}}$ edge of atom W. The sample is probed by a monochromatic x-ray, and the resonant fluorescence is measured in the reflection direction by a CCD and the inelastic fluorescence signals are collected by an energy-resolved fluorescence detector. The distance between collimator and sample surface is 31.0 mm, and the hole diameter and the length of the collimator are 2.8 mm and 20.1 mm. An example of full range fluorescence spectrum is shown in inset at top-left side, and the grey region corresponds to the fluorescence photon energy of L$_\alpha$ line. The inset at top-right side is the reflectivity curve with an incident energy detuning 30 eV from $E_0$, and the pink solid bar indicates the critical angle of Pt (0.46 degree). (b) The values of Re($\eta$) and Im($\eta$) as a function of incident angle, which is calculated by a transfer matrix formulism. (c) The simplified energy levels of W. The driving is labeled by the blue arrow, and the cavity enhanced emission is labeled by the red thick arrow. The inelastic fluorescence decay is labeled by the red thin arrow. }
\end{figure}

Fig. 1(a) depicts the cavity structure used in the present work. The thin-film cavity is made of a multilayer of Pt and C. The top and bottom layers of Pt with a high electron density are used as mirrors. The layers of C in the middle with a low electron density are used to guide the x-ray and to stack the cavity space. In this design, at certain incident angles $\theta_{\textrm{th}}$ below the critical angle of Pt, x-ray can resonantly excite specific cavity guided modes where dips in the reflectivity curve appear as shown in the top-right inset of Fig. 1(a). In the present work, $\theta_{\textrm{th}}$ are $\theta_{\textrm{1st}}$=0.218$^\circ$, $\theta_{\textrm{3rd}}$=0.312$^\circ$ and $\theta_{\textrm{5th}}$=0.440$^\circ$ for the 1$^{\textrm{st}}$, 3$^{\textrm{rd}}$ and 5$^{\textrm{th}}$ odd orders of cavity mode. Then the coupling between the cavity and atom is built by embedding a thin layer of WSi$_2$ at the middle of the cavity where the cavity field amplitudes are the strongest. The field distributions of the 1$^{\textrm{st}}$, 3$^{\textrm{rd}}$ and 5$^{\textrm{th}}$ orders of cavity mode are sketched in Fig. 1(a).

As shown in the middle-right inset of Fig. 1(a), the inner shell energy-level system is different from the simple two-level one, and both resonant channel and incoherent processes such like inelastic radiative channels (Auger decay channels is not exhibited here) can annihilate the core vacancy state, so the decay width is determined by the total decay rates of all relaxation channels. Excited by the incoming x-ray field, the atomic dipole emits the resonant fluorescence through the resonant channel, and the resonant response could be written as a simple form of Lorentz function,

\begin{equation}
f=-f_0\frac{i\gamma_\textrm{re}/2}{\delta+i(\gamma_\textrm{re}/2+\gamma_\textrm{in}/2)}
\end{equation}The electronic continuum in higher energy range is not considered here. $f_0$ is a constant, and $\delta$ is the energy detuning between the incident x-ray energy $E$ and the white line transition energy $E_0$. $\gamma_\textrm{re}$ is the natural spontaneous emission rate of the resonant channel, while $\gamma_\textrm{in}$ is the incoherent decay rate which sums two branches: the radiative decay rate of inelastic channels $\gamma_\textrm{ie}$ and the non-radiative decay rate of Auger process $\gamma_\textrm{A}$, i.e., $\gamma_\textrm{in}=\gamma_\textrm{ie}+\gamma_\textrm{A}$. It is clear that the inverse core-hole lifetime is expressed by the natural width as $\gamma=\gamma_\textrm{re}+\gamma_\textrm{in}$.

Cavity strengthens the photonic density of states~\cite{PhysRevA.51.2545,PhysRevLett.95.097601} at the position of the radiating atom, so the resonant fluorescence will be enhanced. Applying the transfer matrix combined with a perturbation expansion method (SM Sec. I), the resonant fluorescence in the reflection direction is solved as,

\begin{equation}
{{r}_{a}}=-\frac{id{{f}_{0}}\times {{\left| {{a}^{z_a}} \right|}^{2}}{{\gamma }_{\text{re}}}/2}{\delta +{{\delta }_{c}}+i\left( {{\gamma }_{c}}+\gamma  \right)/2}
\end{equation}$d$ is the thickness of the atomic layer, and ${{\left| {{a}^{z_a}} \right|}^{2}}$ is the field intensity where the atom locates. It can be seen that Eq. (2) still has a Lorentzian resonant response, while contains additional cavity effects: the cavity enhanced emission rate $\gamma_c$ and the cavity induced energy shift $\delta_c$,

\begin{equation}
\begin{array}{lll}
  {{\gamma }_{c}}&=d{{f}_{0}}\gamma_{\textrm{re}}\times \operatorname{Re}\left( \eta  \right) \\
 {{\delta }_{c}}&=d{{f}_{0}}\gamma_{\textrm{re}}\times \operatorname{Im}\left( \eta  \right) \\
 \eta &=pq \\
\end{array}
\end{equation}Thus the emission rate is enhanced by a factor of Re($\eta$), where $p$ and $q$ are the field amplitudes corresponding to the wave scattered from up (down) direction into both up and down directions at the position of atomic layer (Sec. I of SM). Note here that the photonic density of states is directly related to the cavity field amplitudes~\cite{PhysRevLett.95.097601,rohlsberger2012electromagnetically}, so Eq. (3) conforms to the typical cavity Purcell effect~\cite{RevModPhys.73.565} which describes the well-known linear relation between lifetime shortening and photonic density of states strengthening. It is clear that the real part of $\eta$ is an essential factor to control the enhanced emission rate, and the energy shift is modified by the image part of $\eta$. The real and image parts of $\eta$ as a function of incident angle are depicted in Fig. 1(b), and $\gamma_c$ and $\delta_c$ are simultaneously modified by the incident angle around the mode angles, which has been observed by Haber \emph{et al} recently~\cite{Haber2019}. On the other hand, Fig. 1(b) suggests that the strongest enhanced emission rate can be achieved without introducing additional energy shift by exactly choosing the angles of odd orders of cavity mode, which will be more convenient to study the individual influence of the CHL on core-hole dynamics (SM Sec. IV). The fully controllable resonant channel makes an adjustable total inverse core-hole lifetime,

\begin{equation}
{{\Gamma }_{n}}=\gamma_c+\gamma_{\textrm{re}}+\gamma_{\textrm{ie}}+\gamma_{\textrm{A}}
\end{equation}where all four contributions are included, herein $\gamma_c$ is the cavity enhanced emission rate, and $\gamma=\gamma_\textrm{re}+\gamma_\textrm{ie}+\gamma_\textrm{A}$ is the natural inverse CHL as the sum of three branches: the natural spontaneous emission rate of the resonant fluorescence channel, the radiative decay rate of inelastic fluorescence channels and the Auger decay rate. $\gamma$ is a fixed value which can be obtained from the experimental spectrum at a large incident angle (Fig. S3 of SM), i.e., $\gamma/2$=3.6 eV. As long as $\gamma_c$ is large enough, this controllable part will dominate the CHL.

As shown in the simplified energy levels in Fig. 1(c), the core-hole lifetime determines the linewidth of inelastic scattering, i.e, the fluorescence spectrum. We employ a RXS formalism known as Kramers-Heisenberg equation to character the inelastic scattering~\cite{gel1999resonant,RevModPhys.83.705} as,
\begin{equation}
{{F}_{if}}\left( \overset{\scriptscriptstyle\rightharpoonup}{k},{\overset{\scriptscriptstyle\rightharpoonup}{k}}',\omega ,{\omega }' \right)=\frac{\left\langle  f \right|{\hat{D}}'\left| n \right\rangle \left\langle  n \right|\hat{D}\left| i \right\rangle }{\delta +i{{\Gamma }_{n}}/2}
\end{equation}Herein the initial state $\left| i \right\rangle =\left| g,\overset{\scriptscriptstyle\rightharpoonup}{k} \right\rangle$, the final state $\left| f \right\rangle =\left| f,{\overset{\scriptscriptstyle\rightharpoonup}{k}}' \right\rangle$, and the intermediate state $\left| n \right\rangle =\left| e,0 \right\rangle$. $\overset{\scriptscriptstyle\rightharpoonup}{k}$ is the wave vector and $\hat{D}$ is the transition operator. For the present system, an intermediate state and a final state are considered, since we choose the L$_{\textrm{III}}$ white line transition and measure L$_\alpha$ with energy $E'$. Eq. (5) indicates that CHL changes can be monitored by the inelastic fluorescence spectrum, and in the present work L$_\alpha$ line (L$_{\alpha1}$ is much stronger than L$_{\alpha2}$) is chosen to obtain the inelastic fluorescence spectra.

%\section{Results and Discussion}
The measurement was performed on the B16 Test beamline in Diamond Light Source. Monochromatic x-ray from a double crystal monochromator was used to scan the incident x-ray energy, and two small apart slits were used to obtain a good collimation beam with a small vertical beam size of about 50 $\mu$m. The multilayer was deposited onto a polished silicon wafer (100) using DC magnetron sputtering method which is popular to fabricate diverse cavity structures~\cite{Rohlsberger1248,rohlsberger2012electromagnetically,PhysRevLett.111.073601,PhysRevLett.114.207401,PhysRevLett.114.203601,haber2017rabi,Haber2019}. Before sample fabrication, the deposition rate was calibrated carefully to guarantee the layer thickness with a good accuracy of better than 1 $\textrm{\AA}$. The size of the wafer is 30$\times$30 mm$^2$ which is larger than the footprint to avoid the beam overpassing the sample length. As shown in the top-right inset of Fig. 1(a), the $\theta-2\theta$ rocking curve with an incident energy detuning 30 eV from the white line position was measured firstly to find the desired specific incident angles corresponding to the 1$^{\textrm{st}}$, 3$^{\textrm{rd}}$ and 5$^{\textrm{th}}$ orders of the guided modes, i.e., the corresponding reflection dips. For a given incident angle, the incident energy $E$ was scanned from 10161 eV to 10261 eV across the transition energy $E_0$=10208 eV. Then the reflectivity corresponding to the resonant channel was measured by a CCD detector, and the inelastic fluorescence lines were measured simultaneously by a silicon drift detector (\emph{Vortex}) with a resolution of about 180 eV at a perpendicular direction. In front of the fluorescence detector, a collimator guarantees a constant detected area of the sample, and the footprint $bw/\textrm{sin}\theta$ on the sample surface is determined by the beam width $bw$ and the incident angle, so the inelastic fluorescence intensities need to be normalized by taking into account a geometry factor~\cite{li2012geometrical}. The strongest L$_\alpha$ fluorescence lines (L$_{\alpha1}$ at 8398 eV and L$_{\alpha2}$ at 8335 eV) of W are far from the white line (10208 eV) and other weak fluorescence lines (9951 eV of L$_{\beta2}$, 7387 eV of L$_l$ and other negligible lines), so L$_\alpha$ can be extracted separately from the energy-resolved fluorescence spectrum.

Firstly, the 1$^{\textrm{st}}$, 3$^{\textrm{rd}}$ and 5$^{\textrm{th}}$ orders are exactly chosen to control the CHL without introducing additional energy shift, and the results are depicted in Fig. (2). Fig. 2(a) shows the experimental and theoretically fitted reflectivity curves. The present theoretical model for resonant fluorescence does not take into account the influence of the absorption edge due to its nature of the electronic continuum, and the continuum overlaps with the right side of the white line. The sudden increase of the absorption coefficient changes the refractive index and dramatically alters the cavity properties~\cite{Haber2019}. So the data below 10220 eV are selected for fitting (labeled by the green region). Above 10220 eV, the theoretical results diverge from the experimental datum. The reflection coefficient includes the contributions from two pathways (SM Sec. II): the first one of $r_0$ is from the multilayer cavity itself that the photon does not interact with the resonant atom, and the second one of $r_a$ is from the resonant atom inside the cavity, i.e., the resonant fluorescence. The linewidth of the cavity is much larger than the one of atom, which means that $r_0$ is more like a flat continuum state and $r_a$ is more like a sharp discrete state~\cite{PhysRevLett.91.193203,PhysRevLett.114.207401}. Therefore, the reflectivity spectrum is a result of Fano interference. It can be seen from Fig. 2(a) that the profile of the reflectivity spectra shows Fano line-shape. The reflectivity spectra give the values of $(\gamma_c+\gamma)/2$ as 7.9 eV, 6.9 eV and 5.2 eV for the 1$^{\textrm{st}}$, 3$^{\textrm{rd}}$ and 5$^{\textrm{th}}$ orders of the cavity mode.

\begin{figure}[htbp]
\centering
\includegraphics[width=0.45\textwidth]{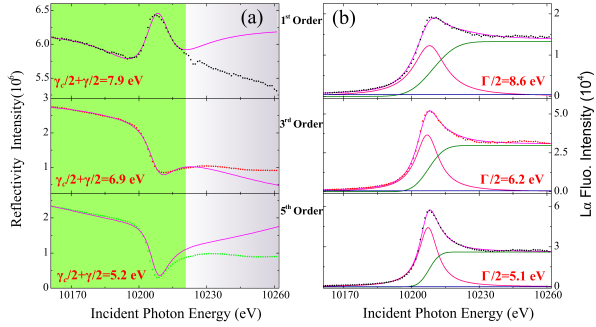}
\renewcommand{\figurename}{Fig.}
\caption{\label{Fig.2}(a) The measured and theoretical reflectivity spectra for the 1$^{\textrm{st}}$, 3$^{\textrm{rd}}$ and 5$^{\textrm{th}}$ orders as a function of incident photon energy. The red solid line is the theoretically fitted result. (b) The measured and fitted inelastic fluorescence spectra of L$_\alpha$ as a function of incident photon energy. The solid lines in pink, red, green and blue are the fitted result, Lorentzian resonance line, electronic continuum line and the flat background respectively.}
\end{figure}

Fig. 2(b) shows the experimental and fitted inelastic fluorescence spectra of L$_\alpha$ as a function of incident x-ray energy for the 1$^{\textrm{st}}$, 3$^{\textrm{rd}}$ and 5$^{\textrm{th}}$ orders. The inelastic fluorescence spectrum is fitted by a custom function combining a simple Lorentz function $L(E)$ with a Heaviside step function $H(E)$ (SM Sec.III), herein $L(E)$ with a linewidth $\Gamma_n/2$ is used to describe Eq. (5) and $H(E)$ is used to describe the absorption edge. The fitted values of $\Gamma_n/2$ are 8.6 eV, 6.2 eV and 5.1 eV which match well with the derived values from the resonant fluorescence spectra of Fig. 2(a), demonstrating that the shortening of CHL indeed comes from the regulation of resonant fluorescence channel. Moreover, the value of $\gamma_c=\Gamma_n-\gamma$ is even larger than $\gamma$ in the 1$^{\textrm{st}}$ order, indicating that the adjustable resonant channel breaks the limitation of Auger processes and unchangeable radiative decay channels and dominantly determines CHL. A behavior of widening linewidth is cross-checked by L$_l$ and L$_{\beta2}$ lines (SM Fig. S5).

\begin{figure}[htbp]
\centering
\includegraphics[width=0.45\textwidth]{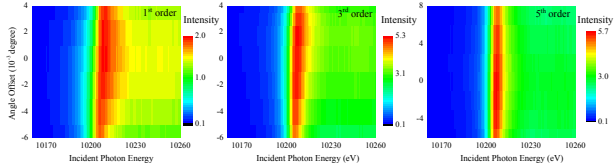}
\renewcommand{\figurename}{Fig.}
\caption{\label{Fig.3}The inelastic fluorescence spectra of L$_\alpha$ as functions of incident photon energy and angle offset. Angle offset is the deviation between the incident angle and the $\theta_{\textrm{1st}}$ ($\theta_{\textrm{3rd}}$, $\theta_{\textrm{5th}}$).}
\end{figure}

The measured inelastic fluorescence 2D spectra are shown in Fig. 3 for selected incident angles around the mode angles of the 1$^{\textrm{st}}$, 3$^{\textrm{rd}}$ and 5$^{\textrm{th}}$ orders ($\theta_{\textrm{1st}}$=0.218$^\circ$, $\theta_{\textrm{3rd}}$=0.312$^\circ$ and $\theta_{\textrm{5th}}$=0.440$^\circ$ respectively). As discussed in Eq. (3), CHL and the cavity induced energy shift are simultaneously controlled by the incident angle. When the incident angle scans across the mode angle, a phenomenon of firstly increasing to maximum at the mode angle then decreasing of the inverse CHL will be observed along with an additional energy shift, which is demonstrated by Fig. 3. For the 3$^{\textrm{rd}}$ order, the maximum linewidth does not seem to be where the angle offset is 0, this may due to the occasionally angle shift from the instabilities of the goniometer or sample holder. Note here that Fig. 3 suggests a way to continuously modify CHL but introduce additional energy shift.

\begin{figure}[htbp]
\centering
\includegraphics[width=0.45\textwidth]{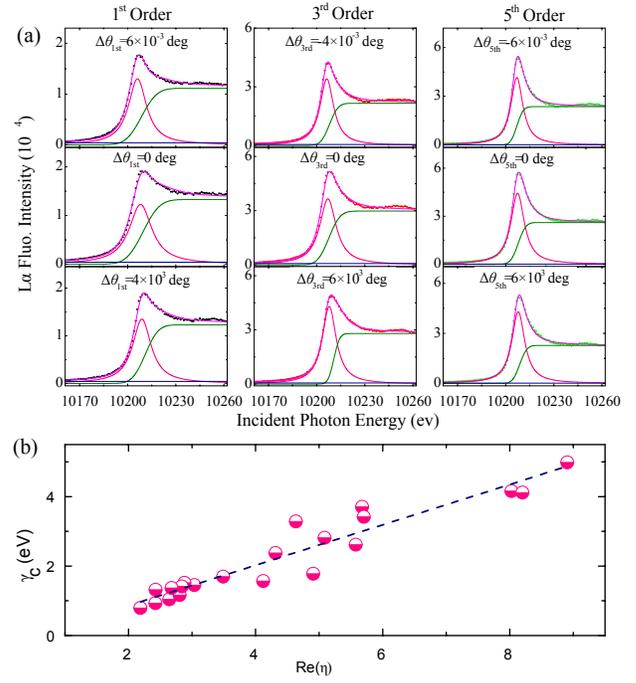}
\renewcommand{\figurename}{Fig.}
\caption{\label{Fig.4}(a) The measured and fitted inelastic fluorescence spectra of L$_\alpha$ for selected angle offsets. (b) Enhanced emission rate $\gamma_c$ as a function of Re$(\eta)$. The values of $\gamma_c$ are derived from the fitted linewidth $\Gamma_n$ as $\gamma_c=\Gamma_n-\gamma$, and the values of Re$(\eta)$ are obtained by transfer matrix calculation. The dashed blue line is a linear fitting of the experimental dots to guide the eyes.}
\end{figure}

As predicted in Eq. (3), the enhanced emission rate $\gamma_c$ is linearly connected with the real part of the cavity filed amplitude $\eta$, and this is the essential to discuss the magnitude of inverse CHL. It should be noted here that the present method to control CHL is different from the scenario of stimulated emission~\cite{PhysRevLett.117.027401,PhysRevLett.121.137403} where a non-linear relationship between the stimulated emission rate and the x-ray field intensity is expected. The present scheme actually employs a cavity to manipulate the enhanced spontaneous emission whose decay rate is linearly determined by the photonic density of states. The inelastic fluorescence spectra in Fig. 3 are fitted to get the values of $\Gamma_n$, and some selected spectra are shown in Fig. 4(a). Then the values of $\gamma_c$ are obtained based on Eq. (4), and the values of Re$(\eta)$ are calculated by the transfer matrix formulism. A good linear relationship between $\gamma_c$ and Re$(\eta)$ is depicted in Fig. 4(b) which is consistent with the prediction of Eq. (3).

From the general viewpoint of cavity-QED in optical regime, the inelastic channel is an incoherent process which accelerates the decoherence, so it is regarded as a defect for the system~\cite{van2013photon}. However, the inelastic channel is a natural character and widely exists in atomic inner-shell systems, herein we demonstrate it can be very useful to monitor CHL changes, enriching the picture of cavity effect.

%\section{Conclusion}
In conclusion, the core-hole lifetime for 2$p$ state of W is manipulated experimentally through constructing an x-ray thin-film planar cavity system. The core-hole lifetime directly depends on the cavity field amplitude at the position of W atom (SM Sec. I and \cite{Huang2020}), which can be adjusted by choosing the different orders of cavity mode or varying the incident angle offset. With a high quality cavity sample, the core-hole lifetime is conveniently manipulated in experiment. Notably for the case of the 1$^{\textrm{st}}$ order, the decay rate of the resonant channel is even stronger than the natural inverse core-hole lifetime which is dominated by the Auger process for L$_{\textrm{III}}$ shell of atom W in common scenarios. Moreover, the inelastic fluorescence spectra are utilized as a good monitor to reflect the core-hole lifetime changes. The cavity structure is suitable for a wide range of x-ray energy from few to tens of keV, so the present scheme could be extended to a lot of elements which have resonant fluorescence channel. Utilizing the present cavity technique, the duration time of RXS process can be controlled not only by the energy detuning, but also by the core-hole lifetime, which will enrich the physical studies for RXS (SM Sec. VI) in future. Combing with the high-resolution $\sim$100 meV analyzer~\cite{hill20072}, a cavity-manipulating RXS is expected to be achievable.

%\section*{\label{Acknowledgements}Acknowledgements}

This work is supported by National Natural Science Foundation of China (Grants No. U1932207), and the National Key Research and Development Program of China (Grants No. 2017YFA0303500 and 2017YFA0402300). The experiment was carried out in instrument B16 of Diamond Light Source Ltd (No. MM21446-1), United Kingdom. Authors thank Xiao-Jing Liu for fruitful discussion.

%See Supplemental Material at http://link.aps.org/supplemental/

%\bibliography{bib}

\end{document}